\documentclass[10pt,twocolumn,letterpaper]{article}

\usepackage[pagenumbers]{arxiv}

\usepackage[linesnumbered,ruled,vlined]{algorithm2e}
\usepackage{graphicx}
\usepackage{amsmath}
\usepackage{amssymb}
\usepackage{booktabs}
\usepackage{gensymb}
\usepackage{hyphenat}
\usepackage[inline]{enumitem}
\usepackage[breaklinks,colorlinks]{hyperref}
\usepackage{multirow}

\usepackage[capitalize]{cleveref}

\begin{document}

\title{Mobile Access Control System Based on RFID Tags and Facial Information}

\author{Kostiantyn Khabarlak \\
Dnipro University of Technology\\
Ukraine\\
{\tt\small habarlack@gmail.com}
\and Larysa Koriashkina\\
Dnipro University of Technology\\
Ukraine
}

\maketitle

\begin{abstract}Better access control system security comes at a higher price. It many cases the price is too high for small companies, leaving them vulnerable with cheap and insecure systems. In this work we introduce an alternative access control scheme, which improves access control security while lowering the cost. In the proposed model, passive RFID tags are mounted near a turnstile or a smart door. Tag reading and programming is done via NFC chip directly on the users smartphone. To enhance security, together with smartphone-based authorization we require the user to provide his photograph while entering a secure gate. The photograph is then displayed on a monitoring dashboard side-by-side with the registration picture, so that the two can be matched against each other. The developed client-server application offers administrative system used to configure gate access policies and monitor entrances with filters by access time, user and gate. Also, we propose a mobile application that allows gate registration and serves as a door unlock key. The suggested access control model reduces installation costs required, while maintaining good security. The system is fully wireless and uses cheap autonomous RFID-tags as its main component. We hope, that the proposed system architecture will find its application in small to medium-sized companies.
\end{abstract}

\keywords{Access Control System, RFID Tags, NFC, Mobile Access Control, Security, Person Identification.}

\blfootnote{
  This article has been published in the peer-reviewed journal. Citation: \nohyphens{K.~Khabarlak, L.~Koriashkina. \emph{Mobile Access Control System Based on RFID Tags and Facial Information}. Bulletin of National Technical University ``KhPI''. Series: System Analysis, Control and Information Technologies, vol.~2, no.~4, pp.~69--74, 2020.} \href{https://doi.org/10.20998/2079-0023.2020.02.12}{DOI:~10.20998/2079-0023.2020.02.12}
}

\section{Introduction}

Many of the modern enterprises use turnstiles or smart doors with access card scanners, where RFID cards are used predominantly. To provide extra security guarantees schools and universities also employ such systems as they are cheap and easy to use. In the same time, such systems have a serious drawback, namely the card can be easily lost, which means an intruder can access the enterprise unnoticed. This in turn may cause critical consequences such as an accident, sensitive information loss, \etc. Installing video surveillance cameras may be a partial solution, which enables detection of an intruder in retrospect. However, storing video surveillance data for a long time may take a lot of disk space. The most efficient yet expensive approach to solving the problem is an installation of expensive biometric systems recognizing face or fingerprint (the latter can be recognized via the terminal or directly on a special access control card).

Using smartphone's NFC chip for secure authentication sees an ever-increasing interest. In this paper, we describe a novel access control scheme, which doesn't use card scanner and offers higher security guarantees when entering a gateway without needing video surveillance cameras.

\section{Review of existing approaches}

To begin with, let us describe existing commercial systems, which control the access by the means of a personal identifier. Company~\cite{KievAC} proposes systems based on using plastic cards or fingerprint. Manufacturer~\cite{PERCo} features a more advanced set of products including virtual mobile cards (NFC-based), bank credit card authentication or biometric systems (fingerprint or facial recognition). A comprehensive list of currently available commercial products is presented in~\cite{HabrCommercialProdList}. These can be categorized in
\begin{enumerate}
  \item Products supporting only classical plastic card id.
  \item Products that additionally include support for NFC or Bluetooth Low Energy (if NFC is not available).
  \item Biometric systems.
\end{enumerate}

Unique identifier in NFC-compatible systems (point 2 above) is granted either via a global server for all clients (in this case a regular fee is taken) or for free based on a unique id of a smartphone (IMEI) or a SIM-card (IMSI). Identifier can be blocked if requested. Wherein, there may arise at least two cases of unauthorized access to the enterprise, that are impossible to track down:
\begin{enumerate*}[label={\arabic*)}]
  \item after having lost the mobile device and before locking its id;
  \item in case of intentional transfer of a smartphone to third parties.
\end{enumerate*}
That is to say that such systems are quite vulnerable on its own.

Let us also highlight some of the more advanced systems. In~\cite{MultiTechPlasticCards} authors note a growing interest in access cards that have an extra level of security. For instance, multi-technology cards that offer embedded fingerprint scanner together with a standard passive RFID tag. To supply the scanner, these cards also have an ultra slim battery. Surely, this comes with a higher price.

Next, let's consider research of promising combined RFID and biometric systems. Patent~\cite{RfidPlusFaceRecognitionPatent} contains a description of a biometric system, in which RFID tag holders are also verified via a standalone facial recognition system. This allows to solve additional problems of access control systems like:
\begin{enumerate}
  \item Buddy badging, when one person logs two badges, while only one actually enters the gate.
  \item Tailgating, when several people enter while using the same badge.
\end{enumerate}
Let any access violation occur, the door will be locked and a special lighting stack will alert the guards to intrude. A similar system with a different alerting method is proposed by~\cite{UniversityHotelAccessControl} for access control in university hostels.

To sum it up, all of the above-mentioned systems have either almost no defense against card transfer (\eg, classical or NFC-based systems) or have a high price (\eg, biometric systems including combined systems).

\section{Mobile access control system model}

\begin{figure*}
  \centering
  \begin{subfigure}{0.49\linewidth}
    \centering
    \includegraphics[width=0.9\linewidth]{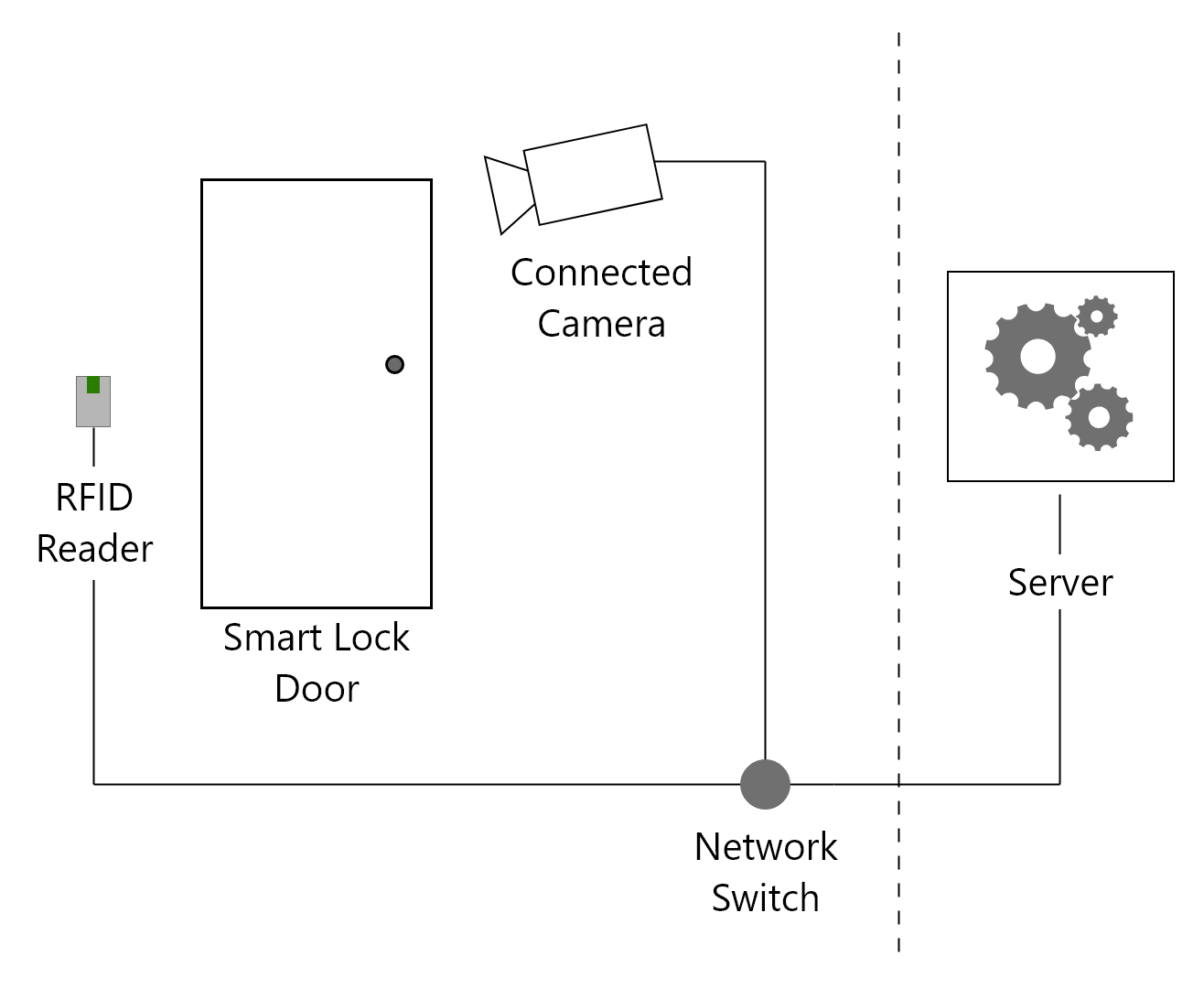}
    \caption{Typical access control system}\label{fig:legacy-control-scheme}
  \end{subfigure}
  \begin{subfigure}{0.49\linewidth}
    \centering
    \includegraphics[width=0.9\linewidth]{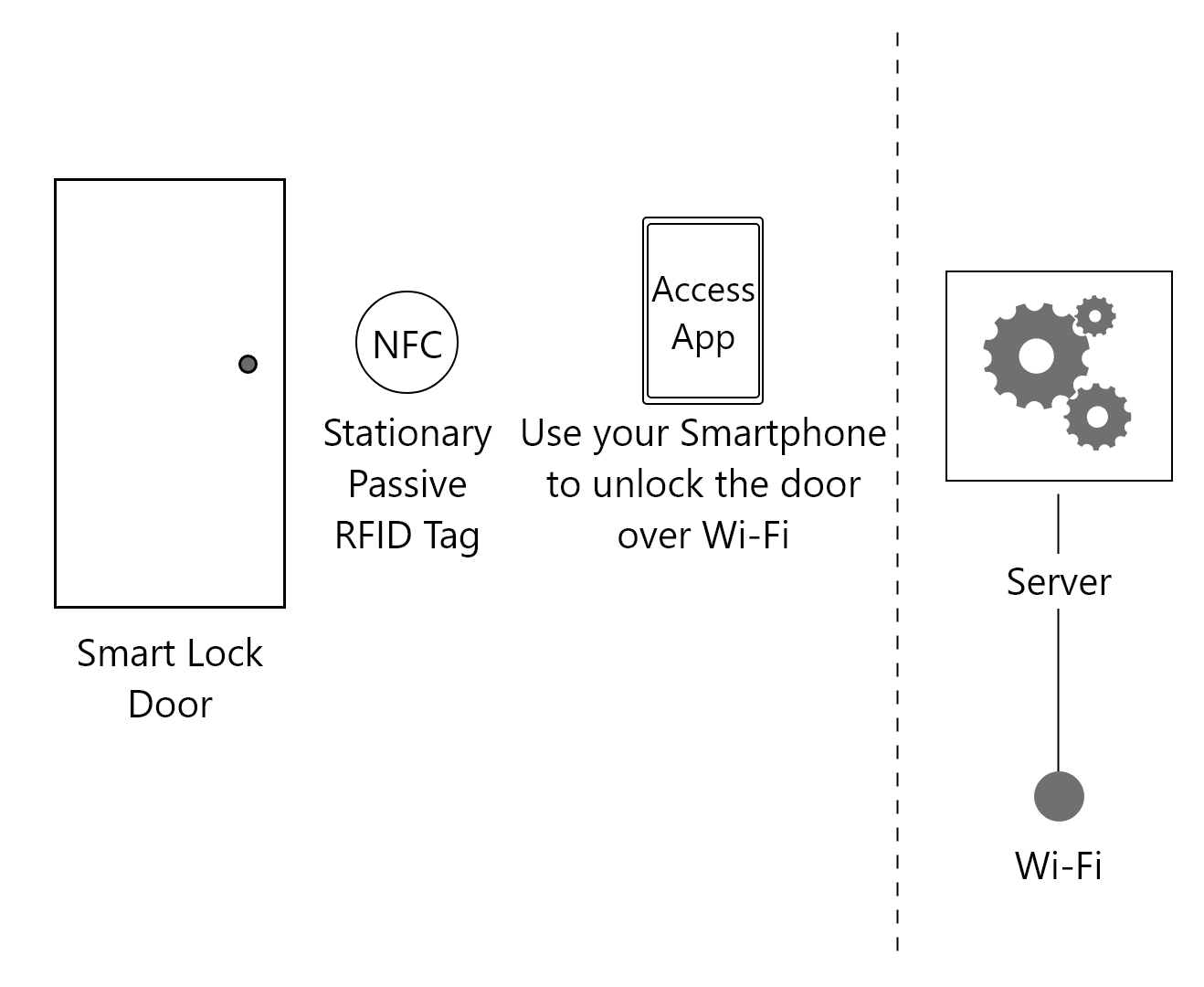}
    \caption{Our system}\label{fig:proposed-control-scheme}
  \end{subfigure}
  \caption{Access control system schemes comparison. In contrast to the conventional system~(a), the proposed system~(b) doesn't require standalone camera or wires to connect to the door. Passive RFID tag is used instead of RFID reader. All communication is performed via the user's smartphone.}\label{fig:control-schemes}
\end{figure*}

In this work we propose to turn the classical access control scheme ``upside-down''. Firstly, instead of a RFID card scanner, which has to be connected to a computer, we propose using passive RFID tags similar to those found in today's plastic cards. The tag will placed near the door and will store the door id and some extra information. RFID tags can store enough data for our system and are much cheaper. Secondly, instead of plastic cards we suggest employing user's smartphone. By holding the device near the passive tag, the proposed access control application will be automatically started. All information about gate location will read from the tag. Also, to avoid the need of a standalone video surveillance camera installation, we require the user to take a photograph on his frontal camera. After that information from the tag joined with the user data is sent to the server via a corporate Wi-Fi network. The information includes: gate information, user id and photograph. In \cref{fig:control-schemes} we present a comparison of typical access control system (shown in \cref{fig:legacy-control-scheme}) and the one we propose (\cref{fig:proposed-control-scheme}). As can be seen from the figure, the proposed system doesn't need camera or RFID scanner installation. Furthermore, no wiring is required as all of the communication is done using smartphone's Wi-Fi connection.

\subsection{Adding tags to the system}

As is known, RFID (Radio Frequency Identification) tag is a device that can store a small amount of data, usually below 888 bytes (while there exist modifications with higher memory capacity, they are rate). The tags are classified into active and passive. Active tags contain an embedded energy source (battery). Their advantage is high acting distance (up to 100m). Passive tags are used in intercoms, biometric passports, contactless credit cards and classical access control systems. Such tags are cheaper than active, but can work only on a short distance ranging from several centimeters to meters depending on a standard and working frequency. As in passive tags the microchip has no built-in power source, an electromagnetic coil is installed instead. A device for reading and programming the tag (including a smartphone) creates electromagnetic field inducting a current in the tag by the Faraday's law~\cite{UniversityHotelAccessControl}.

Smartphones contain the so-called NFC chip (Near Field Communication) to read and program RFID tags. It should be noted, that not all of the existing tags on the market are compatible with NFC. Three main tag types, supporting NFC include: MIFARE Classic~\textregistered, MIFARE Ultralight~\textregistered\ and NTAG~\textregistered. The latter two have the best support among mobile devices~\cite{NfcCompatibility}, and NTAG has the largest capacity. That is why in our product we have decided to use NTAG tags and support two its main modifications: NTAG213 (144 bytes) and NTAG216 (888 bytes)~\cite{NfcTagSpecs}. Besides the writable memory, the tag also contains serial number, an option to enable write password protection and an irreversible switch to read-only mode.

In our system tag programming is supposed to happen on a device of an enterprise security administrator with a use of a special account. The first step is to fill data about the gate to which the tag is going to be attached. The data includes unique gate name and its location. In response to tag registration request, server generates unique unsigned integer id for the gate, which together with the server identifier is written on the tag. In order to avoid data rewrite by third party applications or intentional data corruption, further tag programming is protected via a password (as we have described earlier, this capability is built into the chosen RFID tag standards). The password is global for the given organization and is automatically sent from the server. To complete the gate registration process, the administrator should bring the device to the tag at a distance of 1-3 cm for the programming to take place. Finally, the administrator needs to setup user access policies for the registered tag in a special server-side administrative application (which we will describe in the next section).

The data written to the tag is stored in a special binary format called NDEF (NFC Data Exchange Format). On Android this binary format is implemented via a special NdefMessage message type, containing a set of data records, called NdefRecord~\cite{AndroidNfc}. In our system we write the following information to the tag:
\begin{enumerate}
  \item Server global unique identifier (server GUID), which is used to verify user organization. It should be noted that a distinctive feature of GUID generation is its high randomness, meaning that collisions (generations of the same GUID) are nearly impossible~\cite{Guid}.
  \item Unsigned integer, representing gate id inside the organization.
  \item A special Android Application Record (AAR)~\cite{AndroidNfc} used to launch the application instantly, when the device is held near the tag. The only requirement is that the device should be unlocked..
  \item A similar record for iOS devices, containing the so-called Universal Link.
\end{enumerate}
The calculation of each field's size is shown in \cref{tab:rfid-data}. Overall 120 bytes are written to the tag in our application. Thus, each of the considered RFID tag standards has enough memory for the developed system.

\begin{table*}
  \caption{Data written on the RFID tag.}\label{tab:rfid-data}
  \centering
  \begin{tabular}{lrl}
    \toprule
    Content                    & Size (bytes) & Description                        \\
    \midrule
    Server ID                  & 16           & GUID                               \\
    Secure gate ID             & 4            & Unsigned integer                   \\
    Android Application Record & 42           & Depends on application name length \\
    iOS Universal Link         & 58           & Depends on application name length \\
    Overall                    & 120          &                                    \\
    \bottomrule
  \end{tabular}
\end{table*}

Note, that Apple smartphones did not contain NFC chip for a long time~\cite{NfcCompatibility}. Even after its appearance the use of NFC chip was limited to Apple Pay functionality only. Currently, NFC APIs are being rapidly added to the Apple iOS operating system. Since iOS 11.0 it has become possible to read RFID tags, and iOS 13.0 has introduced a tag write capability~\cite{AppleNfc}. New devices also feature background tag reading support~\cite{AppleBackgroundTagReading}. That is the feature mostly analogous to the AAR. In contrast to Android, the application is not launched automatically, but a notification is presented to the user, inviting to launch it. As we have already mentioned, the iOS launch record is written in Universal Link format~\cite{AppleUniversalLink}. Thereby, all of the functionality required by our system is available on both mobile platforms.

\subsection{Server-side control system}

The main instrument for the company's security administrator is a server-side control panel, implemented in a form of a web-site. Administrative account is needed to access the panel. In there the administrator can add a new user and register its smartphone. To complete the process, users first, last names, and photograph are required. It should be noted, that it is the administrator's responsibility to guarantee the correctness of the entered data. Additionally, gate access policies can be configured in the panel.

\begin{figure*}
  \centering
  \includegraphics[width=0.9\linewidth]{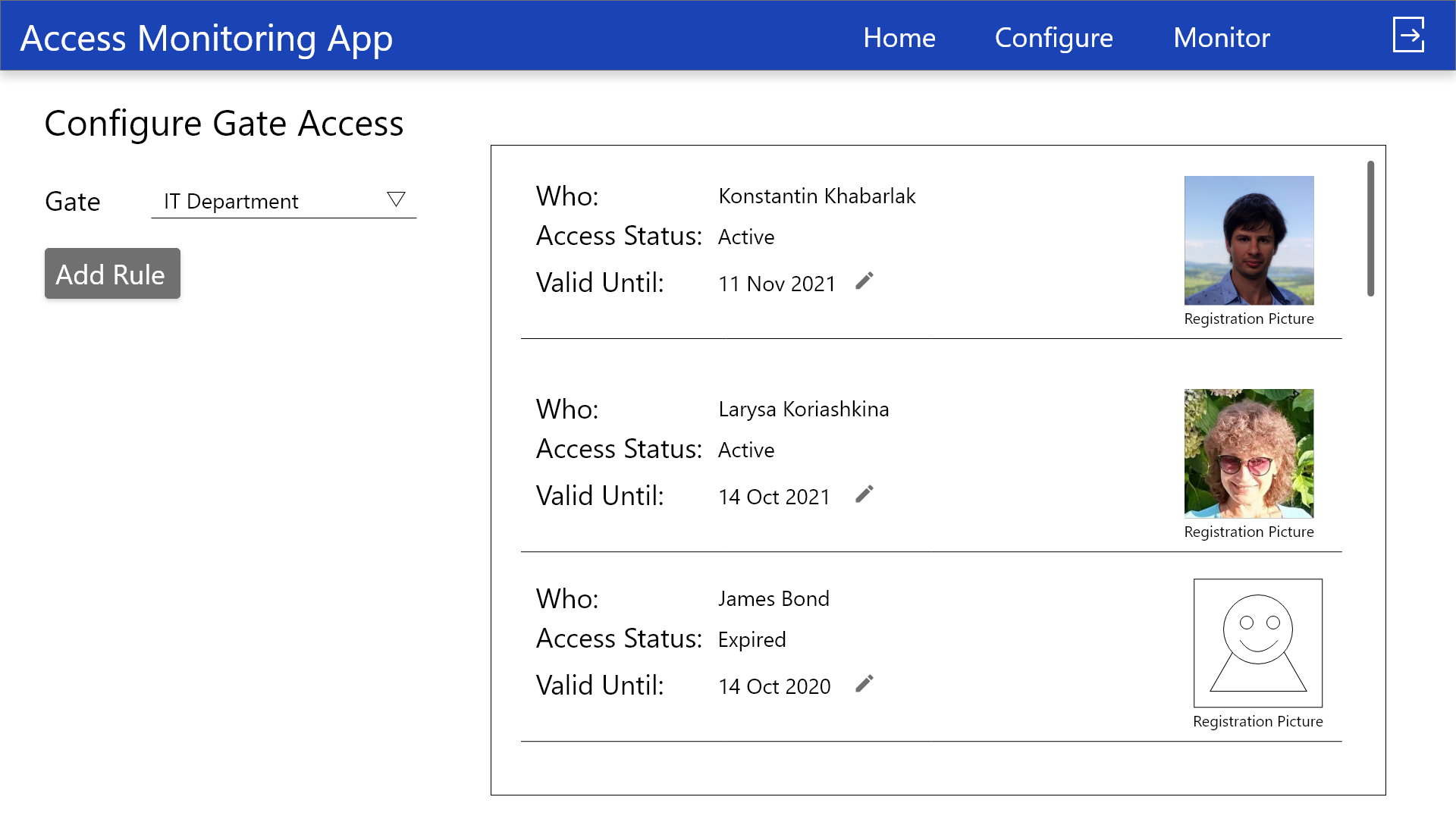}
  \caption{Gate access configuration panel. The gate to be configured is selected on the left. List of users that have (or had) access to the gate are shown on the right. User name, access status, access expiration date and photograph are displayed.}\label{fig:configuration-ui}
\end{figure*}

To setup gate access policies, the tags must be first registered via the mobile application (as it was previously described). Then the administrator can select the tag to configure from a drop-down list. Add, view or edit operations are permitted. Gate access configuration panel is shown in \cref{fig:configuration-ui}.
To enhance the security, each access record should have an expiration date set, after which the access will be automatically disabled.

\begin{figure*}
  \centering
  \includegraphics[width=0.9\linewidth]{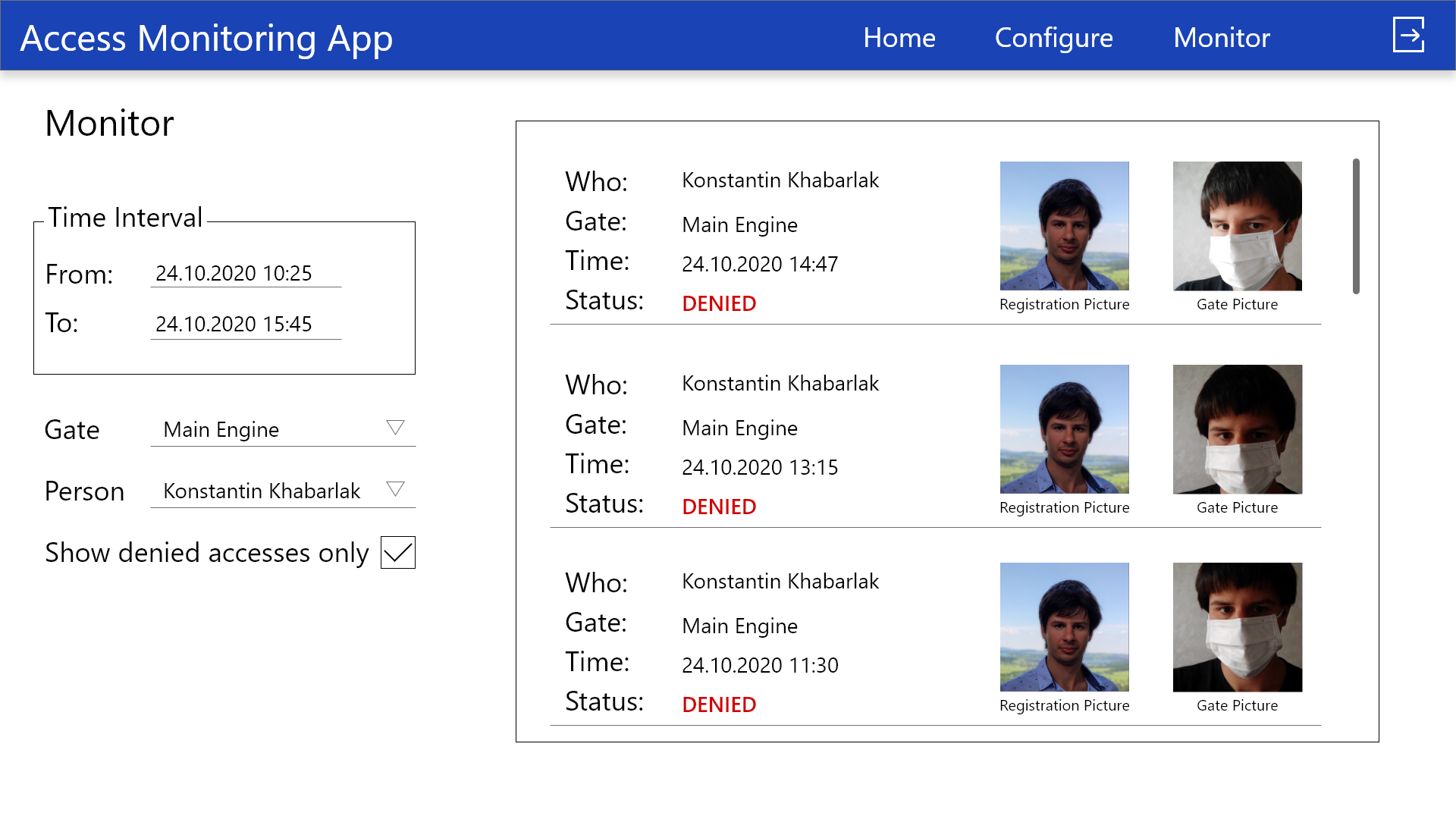}
  \caption{Monitoring tab of the proposed access control system. Access list filters by time, gate, person or access status are configured on the left. Corresponding user accesses are shown on the right. The following information is presented: user's name, registration photo and picture taken at the gate, gate name, access time and status.}\label{fig:monitoring-ui}
\end{figure*}

After the initial setup, the main panel that we expect to be used is the monitoring panel (shown in \cref{fig:monitoring-ui}). A number of monitoring features are proposed:
\begin{enumerate}
  \item An ability to view access records in real time or by time filter (for example, during or outside the working hours).
  \item Filter by the tag to which an access attempt has been made.
  \item Filter by user.
  \item An option to display denied accesses only.
\end{enumerate}
Each of the filters can be left empty if need. In future we suggest extending the system with mobile face recognition based on research presented in~\cite{FastLandmarkDetectionSurvey,FaceDetectionOnMobile}.

\subsection{Backend and third-party services integration}

Along with the above-described user-facing parts, we have a server backend used to communicate with the mobile application via REST API. Also, we have implemented a SignalR~\cite{SignalR} endpoint for third-party services integration. SignalR provides a feed to access control events, and is designed for mobile and web integrations. In our system we do not propose turnstile or smart door systems, so we expect the end-users to be able to quickly adapt their existing systems via the provided API. We hope that the API provided will allow for a seamless implementation of our system into existing infrastructure.

\section{Conclusion}

Access control client-server application has been presented in the paper, which includes:
\begin{enumerate}
  \item Administrative system to configure gate access policies.
  \item Monitoring system with filters by access time, user and RFID-tagged gate.
  \item Mobile application made to register gates and that serves as a key to unlock the doors.
\end{enumerate}

The implementation of the developed system will allow to lower the cost of access control systems in schools, universities and enterprises by replacing stationary RFID scanner with a cheap tag, and also by avoiding video surveillance camera installation. The camera is not required as the user takes a photograph on his mobile phone when unlocking the door.

We hope, that the proposed application will make a contribution to the development of more secure and less expensive access control systems.

We see an implementation of a mobile face recognition system as one of the next steps to enhance the proposed application.

\begin{small}

  \printbibliography

\end{small}

\end{document}